\documentclass[journal]{IEEEtran}
\usepackage{color}

\usepackage{cite}
\usepackage[pdftex]{graphicx}

\usepackage{multirow}
\usepackage{amsmath}
\usepackage{array}
\usepackage{eqnarray}
\usepackage{fixltx2e}
\usepackage{url}
\usepackage[normalem]{ulem}

\begin{document}
%
\title{Evaluating Spintronic Devices Using The Modular Approach}
%
%
%

\author{
\IEEEauthorblockN{Samiran~Ganguly,
        Kerem~Yunus~Camsari,
        and~Supriyo~Datta}\\
\IEEEauthorblockA{School of Electrical and Computer Engineering, Purdue 
University, West Lafayette, IN 47907\\
Emails: \{sganguly, kcamsari, datta\}@purdue.edu}
}
\maketitle

\begin{abstract}
Over the past decade a large family of spintronic devices have been 
proposed as candidates for replacing CMOS  
for future digital logic circuits. Using the recently developed Modular Approach framework, we investigate and identify the 
physical bottlenecks and engineering challenges facing current spintronic devices. 
We then evaluate how systematic advancements in material properties and device 
design innovations impact the performance of spintronic devices, as a possible 
continuation of Moore's Law, even though some of these projections are 
speculative and may require technological breakthroughs. Lastly, we illustrate 
the use of the Modular Approach as an exploratory tool  for probabilistic 
networks, using superparamagnetic magnets as building blocks for such networks. 
These building blocks leverage the inherent physics of stochastic spin-torque 
switching and could provide ultra-compact and efficient hardware for 
beyond-Boolean computational paradigms.
\end{abstract}

\begin{IEEEkeywords}

\end{IEEEkeywords}

%

%
%
%
%


\section{Introduction}

\IEEEPARstart{T}{here} has been enormous progress in the last few decades,
effectively combining spintronics and magnetics into a powerful force that is 
shaping the field of memory devices, while new materials and phenomena continue 
to be discovered at an impressive rate 
\cite{mihai_miron_current-driven_2010,liu_spin-torque_2012,
chen_experimental_2009,heron_deterministic_2014,wang_electric-field-assisted_2012,liu2012giant} 
providing an ever-increasing toolbox for the design of novel functional devices 
\cite{behin-aein_proposal_2010,datta_non-volatile_2012,nikonov_proposal_2011,
zhu_mlogic:_2015, 
manipatruni_spin-orbit_2015,chen_domino-style_2015,roy_hybrid_2011,imre_majority_2006}. It is natural to ask whether these 
developments can be harnessed to meet the increasing interest in finding new 
ways to meet the challenge of continuing the celebrated Moore's law 
in the coming decades.

Broadly speaking the relevant developments in spintronic materials and 
phenomena belong in two categories, those that enable conversion of electrical 
into magnetic information (the WRITE function) and those that enable the 
reverse process (the READ function). READ and WRITE functions are of course 
central to memory devices and it has also been shown that they can be integrated 
into a transistor-like device, with gain and input-output isolation, that we 
call a ``spin switch'' which can be used as a building block for logic circuits 
\cite{datta_non-volatile_2012,datta_what_2014}.

\begin{figure}[!t]
\centering
\includegraphics[width=2.8in]{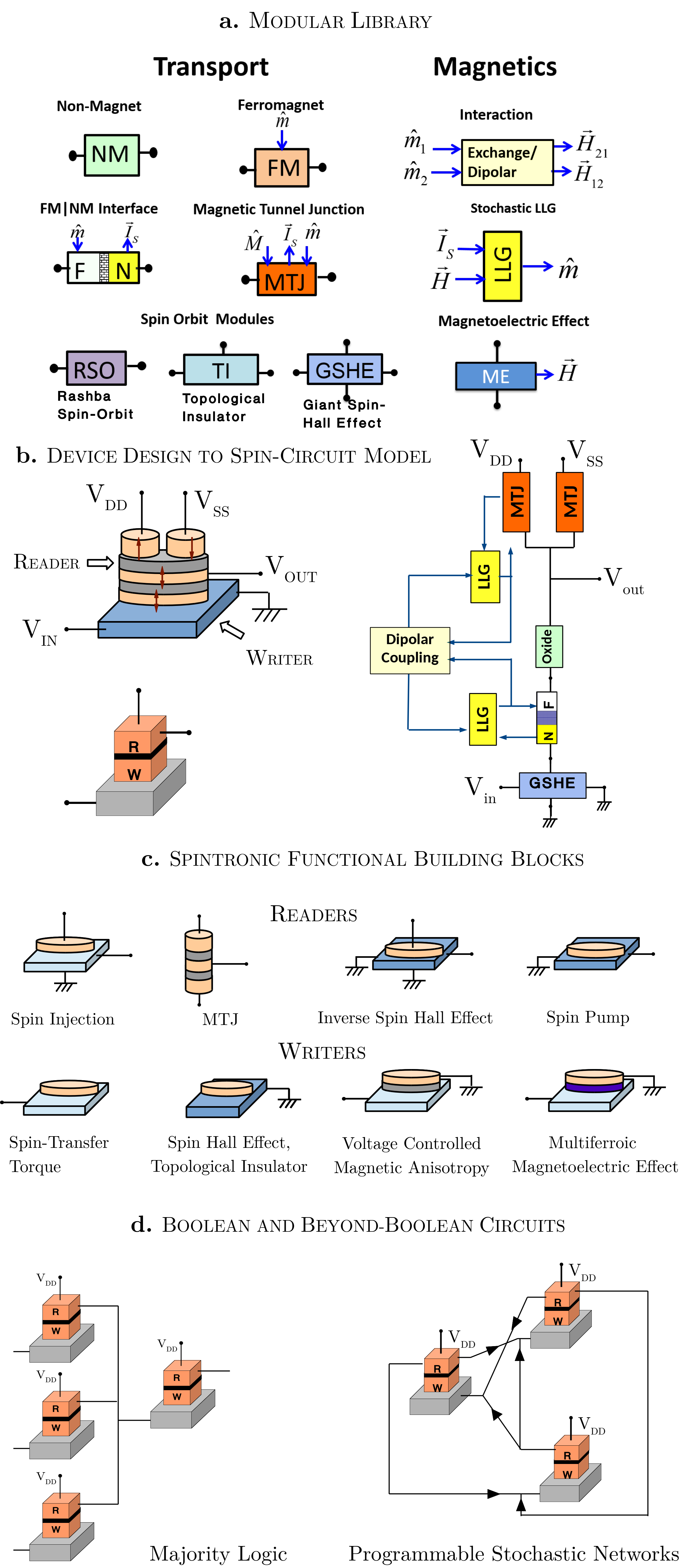}
\caption{\textbf{a. Modular 
Spintronic Library:} Benchmarked modules for charge and spin transport and magnetic phenomena that can be used to create 
spin-circuit models for spintronic devices. For detailed description 
of the library, please see \cite{_modular_web} \textbf{b. Device Design to 
Spin-Circuit Model:} (top left): Spin switch, an example spintronic 
logic device. (bottom left): Compact representation of spin switches used 
throughout this work. (right): Spin-circuit models for a device can be built 
by connecting together various modules, using 4-component currents and voltages 
and simulated in SPICE like programs. \textbf{c. Spintronic Functional Building 
Blocks:} A sampling of various readers and writers that can be used to build 
spintronic devices. \textbf{d. Boolean and beyond-Boolean Circuits:} 
Spin-circuit models can be used to design and evaluate novel, 
ultra-compact, and efficient spintronic-based circuits and architectures. }
\label{fig_intro}
\end{figure}

A natural question to ask is how such a spin switch compares with a standard 
switch based on CMOS (complementary metal oxide semiconductor) technology, and 
several authors have addressed different aspects of this question 
\cite{nikonov_benchmarking_2015,kim_spin-based_2015, manipatruni_material_2016, 
chang_scaling_2016}. The purpose of this paper is to establish a systematic 
framework for evaluating the impact of different READ and WRITE units on the key 
performance criteria for logic devices, namely their static power consumption, 
switching energy, switching delay and the energy-delay product. This framework 
is based on Modular Approach to Spintronics  \cite{camsari_modular_2015,_modular_web} 
whereby different materials and phenomena are represented by experimentally 
benchmarked modules, whose input and output voltages and currents have four 
components, one for charge and three for spin (fig. \ref{fig_intro}a). These 
modules can then be combined using standard SPICE or SPICE-like platforms to 
evaluate circuit and system level performance.

In Section II, we analyze a series of spin switches based on a magnetic tunnel 
junction (MTJ) for the reader and the spin Hall effect (SHE) for the writer 
providing a direct comparison of switching currents and power with a series of 
CMOS inverters (fig. \ref{fig_intro}b). The purpose is to pinpoint the factors 
underlying the inferior static power and energy-delay product of spin switches 
relative to a CMOS switch. In Section III using modular approach we evaluate 
spin switches utilizing 
several alternative readers and writers (fig. \ref{fig_intro}c) which show a 
potential improvement in performance that could be comparable 
CMOS if optimistic material performance and integration parameters are realized 
in the future.

Since Moore's law has been made possible by a doubling of the number 
of switches in a given area every two years, this cannot continue without a 
significant reduction in the energy-delay product relative to CMOS which seems 
difficult based on present state-of-the-art in spintronics. However, it has been 
noted that from a consumer perspective, Moore's law represents a doubling of 
``user value'' every two years and this could be enabled through enhanced 
functionality \cite{waldrop_chips_2016}. It has been recognized that nanomagnets 
could provide enhanced functionality over CMOS through their unique physics 
that provides a natural bistability, threshold response and stochastic 
operation.

We end in Section IV with an evaluation of a simple version of a 
restricted Boltzmann machine featuring stochastic spin switches (fig. 
\ref{fig_intro}d) and is similar to other examples of usage of stochasticity of 
magnets for computing 
\cite{venkatesan_spintastic:_2015,us_department_of_commerce_stochastic_????,
locatelli_spintronic_2015,
khas_physically_2015}. 
Such ``stochastic computers'' are commonly implemented virtually using software 
algorithms on a deterministic hardware platform, but nanomagnets could provide a 
natural physical hardware for their efficient implementation 
\cite{behin-aein_transynapse:_????,sutton2016intrinsic,PSLpaper}. A detailed evaluation of different options 
and possibilities is beyond the scope of this paper. Our purpose here is simply 
to use the Modular Approach to establish a common framework for connecting from 
basic materials and phenomena all the way to circuits and systems, both 
deterministic and stochastic.


\section{Spin-Switches vs. CMOS}

In this section we use the original proposal of the spin switch 
\cite{datta_non-volatile_2012} as an example to examine the physics of 
power dissipation in spintronic devices. The spin switch uses an 
MTJ stack for READ, while WRITE is through a GSHE layer 
driving a ferromagnet (FM) coupled magnetically to the MTJ, 
providing electrical isolation and coupling, while the GSHE-based writing
provides gain. In section III, we systematically go through other variations and 
possibilities of READ/WRITE units that can be used to build  
a family of spin switches. 

\begin{figure}[!t]
\centering
\includegraphics[width=3.2in]{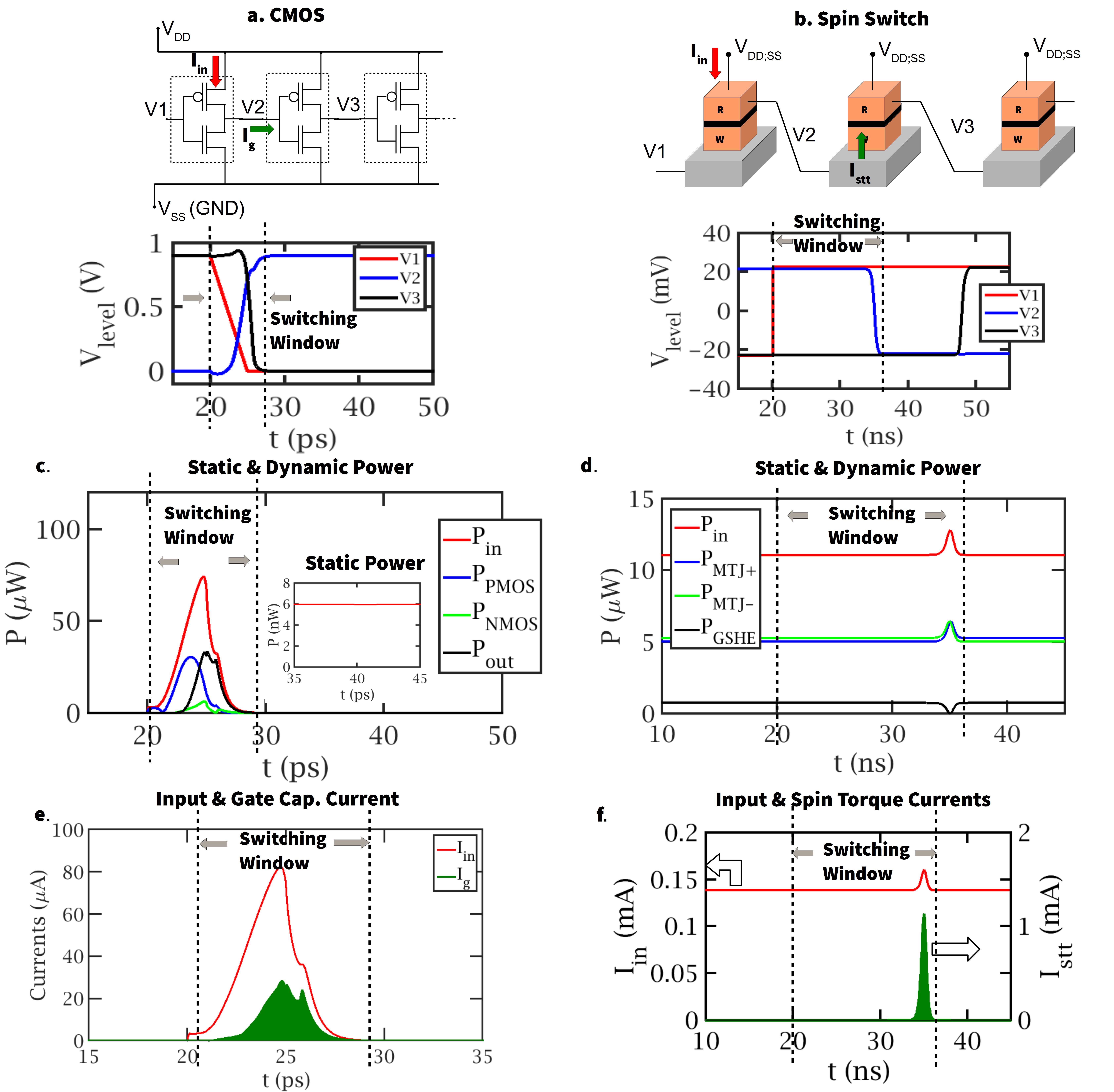}
\caption{\textbf{a. Circuit Testbench for CMOS:} FO-1 Inverter chain built using CMOS inverters. The transient 
simulation shows a switching event. A change in V1 from high to low level cause 
V2 and V3 to change sequentially. \textbf{b. Circuit Testbench for spin switch:} FO-1 Inverter chain built using spin-switch. The transient simulation shows a switching event. A change in V1 from high to low level cause 
V2 and V3 to change sequentially.  \textbf{c. Static and Dynamic Power in CMOS:} Dynamic power dissipation in first CMOS inverter in the chain. Power dissipated  in the PMOS and the NMOS transistors, the output (the next stage), and the total power provided by the supply rail are shown.  \textbf{(Inset)} Power dissipation at the steady state. 
\textbf{d. Static and Dynamic Power in the spin switch:} Static and 
Dynamic power dissipation in the first spin switch. Power dissipated in the two 
MTJs (reader), the output (GSHE of the next stage), and by the total power 
provided by the supply rails are shown. \textbf{e. Switching and Total Charge in 
CMOS:} (Green area) Charge involved in switching the CMOS inverter, i.e. the 
charging of the gates, and (red) total charge supplied by the 
source in the switching process. \textbf{f. Switching and Total Charge in spin switch:} (Green area) Charge involved in switching the spin switch, i.e. the spin-torque switching of the write magnet, and (red) total charge supplied by the 
source in the switching process.} 
\label{fig_devmet}
\end{figure}

As a reference point, we have used a CMOS inverter built from ASU-PTM models for 
14nm HP-FinFETs \cite{_predictive_????,zhao_new_2006} to highlight the 
differences and similarities between physics of switching and power dissipation 
in both charge based and spin based devices. Fig. \ref{fig_devmet}a,b shows a 
FO-1 inverter chain built using CMOS and the spin-switch respectively. Details 
on simulation parameters are shown in the supplementary. For details on spin-circuit modeling 
and to obtain open-source circuit models of the devices used in this work, 
please see \cite{_modular_web}.

\subsection{Static  Dissipation}

Static power dissipation is the Joule heating ($I^2 R$) losses at  steady 
state, i.e. when the device is not switching. It is a critical measure of device performance, 
since this contributes to the thermal budget of any circuit built using the 
device. 

Insets in fig. \ref{fig_devmet}c,d show a typical switching transient and 
toggling of the voltages causally ($V1 \rightarrow V2 \rightarrow V3$) in the 
logic pipeline. The supply voltage levels used in HP-CMOS are typically 
$700-900\ \rm mV$, whereas in spin-switch at minimal overdrive, it can range 
from  
$10-100\ \rm mV$ and is sufficient to generate the 
threshold spin current necessary ($100 - 200\ \rm{\mu A}$) to switch a typical 
nanomagnet with a 40 $k_B T$ stability, given the characteristic resistances of 
all the components (GSHE, MTJ) are around $0.5 - 2\ \rm{k\Omega}$ (A complete 
list of 
parameters  are shown in the supplementary). 

The static power levels in HP-CMOS are of the order of 
$\rm nW$ (fig. \ref{fig_devmet}c) as expected from the leakage current levels 
\cite{_itrs_????}, which is much 
lower compared to dynamic power levels. However, the static power dissipation 
in 
spin-switch is nearly the same as the dynamic power dissipation because the 
spin-switch does not turn off  at  steady state (fig.\ref{fig_devmet}d). There 
is  a constant  current flow through the MTJs, since the resistance ratio 
$R_{AP}/R_{P}$ of the MTJ pair is $3 - 4$, compared to CMOS where 
$R_{OFF}/R_{ON}$ of the PMOS$-$NMOS pair is $10^4 - 10^{5}$. 

Additionally, the input side of the spin-switch (GSHE) is a low impedance 
component, unlike the CMOS where the MOSFET gate terminals are high impedance 
components that  become open circuit  at steady state. As a result of these two 
static current flows, the static power dissipation remains near $\rm \mu 
W$ levels for 
the spin-switch.

\subsection{Dynamic Dissipation: Energy $\times$  Delay }

Any alternative to CMOS  needs to be competitive in terms of both the switching 
delay and switching energy.
A useful measure of dynamic dissipation is the product of the switching energy 
$\times$ switching delay ($E\times\tau$ product) per switching event, because a 
switch can be overdriven for lower switching delay, 
resulting in higher switching energy or vice versa.  It has been shown 
\cite{sarkar_charge-resistance_2014} that the $E\times\tau$ product can be related to 
the charge consumed during the switching process, i.e. $E \times \tau 
= Q_{sw}^2 R$, where $Q_{sw}$ is the charge drawn over the effective 
resistance $R$ to switch the device. The $E\times\tau$ metric recast as $Q_{sw}^2 R$ 
provides a powerful approach to understanding the 
fundamental limits of dynamic dissipation in a device by quantitatively 
relating it to its physical properties and opens a pathway to better 
component design for higher performance.

\subsubsection{Switching Delay}

The timescale of switching in a scaled CMOS inverter, a 14 nm FinFET in this 
case, is of the order of $\approx 1-10 \rm \ ps$ (fig. \ref{fig_devmet}a). The 
 switching delay for magnets is dependent on the current overdrive: 
 At \textit{large overdrives} the  delay can be determined by angular momentum 
considerations, where a total amount of charge that is twice the number of spins 
($M_s\Omega/\mu_B$) needs to be deposited for a complete reversal of 
magnetization. This means that the switching delay inversely scales with the 
driving current, i.e.   $\tau \sim 2 q M_s\Omega/\mu_B I_s$. The 
exact switching delay is a strong function of the initial angle of the 
magnetization \cite{sun_spin-current_2000}. In our simulations, the  initial 
angle is chosen to be the rms of the equilibrium deviation from the easy axis 
\cite{behin-aein_proposal_2010}.  


\subsubsection{Switching Energy and Switching Charge}

Switching energy in itself can be calculated by simply integrating the power 
between the switching window. While this approach does work numerically, to gain 
better physical understanding of dynamic dissipation, we look at the 
switching charge, as discussed at the beginning of the section. To do so from 
our circuit testbench, we first integrate the input current (red arrows indicated in 
fig. \ref{fig_devmet}a,b) supplied from the sources within the switching window, 
whose 
starting point is the the time-point at which the input signal signal starts 
changing and the ending point is the time-point at which the output signal is 
within $1\%$ of its final value. This provides the total charge per switching 
event. Additionally, we integrate the total current deposited to the CMOS input 
terminals (gates) and we do the same for the analogous quantity for magnets, the 
z-component of the spin-torque current ($I_{STT;z}= 
(\hat{m}\times\overrightarrow{I_{s}}\times\hat{m})_{z}$) which are indicated by 
the green arrows in fig.\ref{fig_devmet}b.

We find that the area under the green curve for the CMOS (fig. 
\ref{fig_devmet}e) is about $220\ e^-$, equivalent to the charge 
deposited to the CMOS gates $Q_{sw}=(C_{gate}^{PMOS}+C_{gate}^{NMOS})V_{DD}$, 
whereas the green area for the spin-switch (fig. \ref{fig_devmet}f) is about 
$Q_{sw}= 2 q N_s \approx 1,600,000\ e^-$  where $N_s = M_s\Omega/\mu_{B}$ is 
the total number of spins ($\mu_B)$. Therefore, reducing the dynamic 
dissipation for any logic device involves 
scaling  the  $Q_{sw}$ through better component 
design\cite{behin-aein_switching_2011,camsari_ultrafast_2016}.

It should be noted that the $Q_{sw}$ in our discussion this far is 
only an approximate measure of $E\times\tau$ since it does not incorporate the 
steady state currents discussed in static dissipation section, as well as any 
load driven by the device at the fanout. Using our methodology, we 
can relate the $Q_{sw}$ to the total charge $Q_{T}$ 
provided by the supply.  As an example in fig. \ref{fig_devmet}f, the area 
under the red curve integrated in the switching window, gives the total  charge 
pumped in by both the supplies ($V_{DD}, V_{SS}$) during one switching event. 
This area is about 10 times the green area which gives the $Q_{sw}$, in the 
case of spin-switch. For the  CMOS inverter (fig. \ref{fig_devmet}e), the total 
charge pumped is about 6 times the $Q_{sw}$. Any improvement in scaling down the 
dynamic dissipation will then involve reducing the $Q_{sw}$ as well the $Q_T$ 
during switching.

\begin{table}[!h]
\begin{center}
 \begin{tabular}{|l|l|l|}
\hline
 Metric & HP-CMOS & Spin-Switch\\ \hline
 Voltage Level & $\sim 0.8\ \rm V$ & $\sim 20\ \rm mV$ \\
 Static Power & $\sim 10^{-3} \mu \rm W$ & $\sim 10 \mu \rm W$ \\
 Dynamic Power & $\sim 10^{2} \mu \rm W$ & $\sim 10 \mu \rm W$ \\
 Switching Delay & $\sim 10^{-2}\ \rm ns $ & $\sim 10\ \rm ns $\\
 \textbf{Switching Charge, $\mathbf{Q}$} & $ \sim 10^2 - 10^3\ e^- $ & $\sim 10^5 - 10^6\ e^- $\\
 \textbf{$\mathbf{R_{OFF}}$/$\mathbf{R_{ON}}$} &  $\sim 10^5$   & $\sim 4 $ \\
 \hline
\end{tabular}
\end{center}
\caption{Comparison between HP-CMOS and Spin-Switch Dissipation}
\label{tab_devmet}
\end{table}

We summarize the metrics discussed in this section in Table.~\ref{tab_devmet}. 
While the details of the physics in these examples depend on the specifics of  
circuit design, material parameters and overdrive conditions, we believe that  
the measure of static power in terms of $R_{OFF}/R_{ON}$ (READ) and dynamic 
power in terms of $Q_{sw}$ (WRITE) are  general notions that should be 
applicable to 
any logic device. 


\section{Alternative Designs for Spin-Switch}
In this section, we  show how the spin-switch device can be made competitive 
with CMOS inverter  through improvements in device materials, magnetic stack 
designs, and use of different components based on new phenomena. This allows us 
to project performance enhancements quantitatively and build possible roadmaps
for optimized spin-switch devices. We have covered only a small sampling of 
possible optimization and design space for spintronic devices. Our main purpose 
is to demonstrate the power of Modular Approach in building and evaluating  
alternative designs. 


\begin{figure}[!t]
\centering
\includegraphics[width=3in]{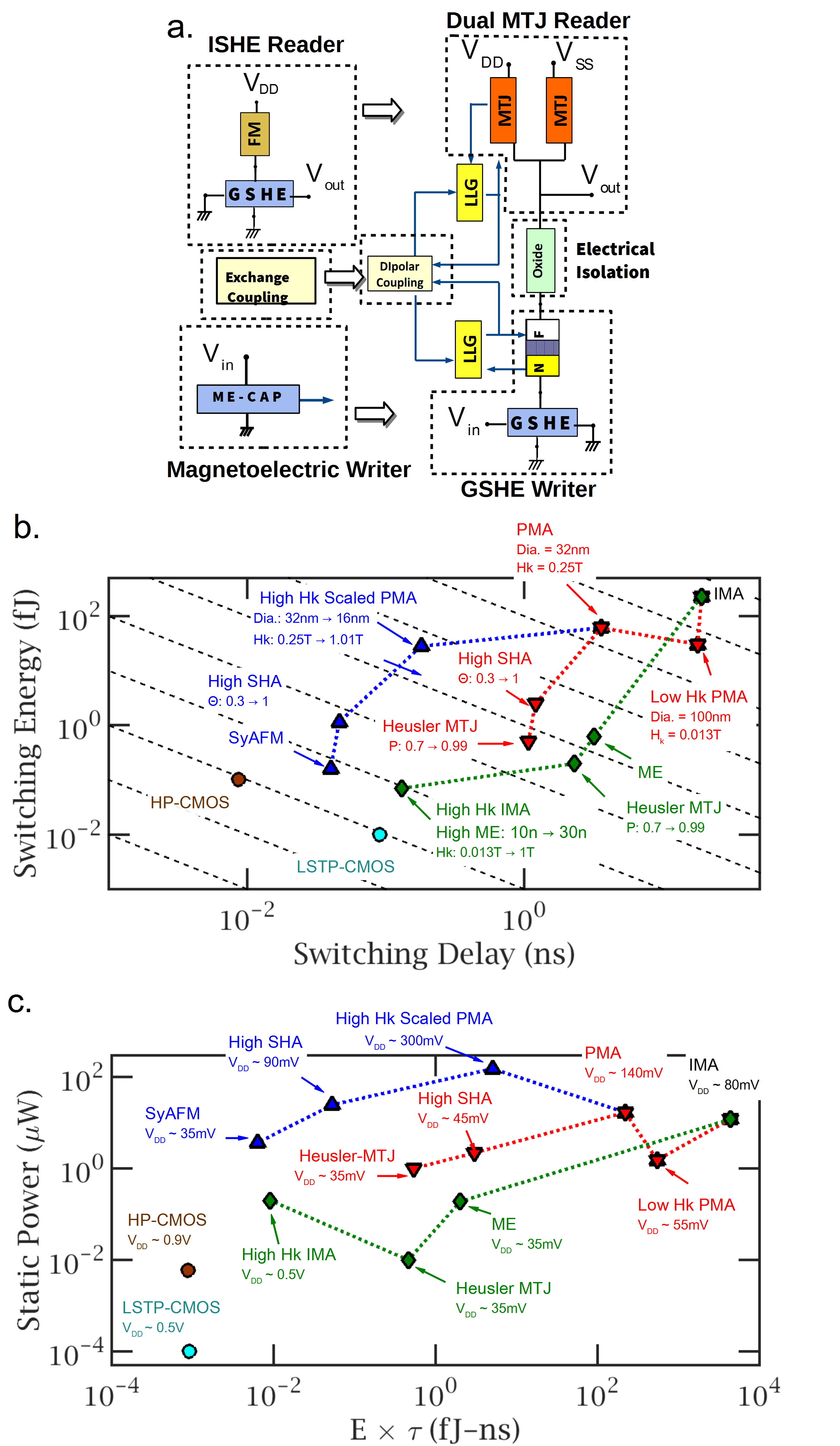}
\caption{\textbf{a. Alternative Designs for spin switch:} Modular Approach to Spintronics
allows us to create alternative designs by replacing an individual module with another, without 
changing the full spin-circuit model. These alternative designs can be evaluated 
for their $E\times \tau$ and static power. The figure illustrates 
modification to the spin switch can be incorporated by 
swapping appropriate modules in the spin-circuit model, due to the 
built-in modularity of the spin switch in terms of READ and WRITE units. 
\textbf{b. $\mathbf{E\times\tau}$ 
Improvement through Material Optimization:} Investigating 
the improvement of $E\times\tau$ under various 
alternative materials and device shown in three different design trajectories, and 
comparison with CMOS. The three trajectories employ innovations such as high 
$H_k$ scaled PMA magnets, Heusler alloys, high spin-Hall angle (SHA) materials, 
magneto-electric effect in multiferroics (ME) , and synthetic ferrimagnets. The 
optimizations applied together can bring the performance of spin switches to 
within an order of magnitude of $E\times\tau$ of both high performance and low 
standby power CMOS (HP-CMOS, LSTP-CMOS). \textbf{c. Static Power Improvement 
through Material Optimization:} Investigating the static energy dissipation for 
the three different design trajectories (same as in \textbf{b.})}
\label{fig_dynamic}
\end{figure}

\subsection{Alternative Design Trajectories}

We show 3 example trajectories of how the spin-switch can be optimized by 
stacking one change on top of the other and observe its effect in reducing 
the static and the dynamic  dissipation. The next few headings describe the 
changes we introduce to the spin switch design sequentially. While each of 
these improvements can be applied independent of each other, the presented 
information shows how the performance of the spin-switch can 
improve dramatically when these optimizations are applied in succession. 

The numbers reported in this section are from the measurements 
of static power, switching energy and switching delay performed for an 
FO-1 circuit testbench, as described in the supplementary. 
These numbers would vary if a different testbench, such as 
FO-4 or 32-bit adder, is chosen, since it will change the loading conditions. 
While some of the low $Q_{sw}$ spin-switch designs could work with ultra low 
voltages ($< 20\ \rm{mV}$), a conventional circuit may not be able to provide 
$V_{DD;SS} < 50 \ \rm{mV}$ due to the inability to deliver 
sufficient power at such low voltages \cite{gatechpowerthesis}. Additional transistors could be used in 
conjunction with each spin-switch to obtain lower voltages and such transistors 
will have their own dissipation. We have used a minimum supply voltage level of 
$35\ mV$ in some of the alternative designs but do not include dissipation 
numbers of transistors in such cases. Furthermore, in this work our analyses 
are limited to the devices themselves and we do not measure dissipation in 
interconnect and supply rails as well as the driver circuitry consisting 
of CMOS devices. All the material parameters used in 
these simulations are listed in the supplementary and are optimistic projections 
which may not have been demonstrated at present. We also list theoretical 
minimum 
operating points for the spin switch designs discussed in this section based 
on analytical model for these devices and the material parameters.

\subsubsection{Magnet Design$-$IMA to PMA}

The first improvement we make to the device design is the replacement of 
magnets with in-plane magnetic anisotropy (IMA) with perpendicular 
magnetic anisotropy  (PMA). This reduces the minimum switching current 
required by getting rid of the large demagnetizing field ($h_d = 4\pi M_s/H_k$) 
while maintaining the barrier height $U$, since in the monodomain approximation, the in-plane minimum switching 
current is given by \cite{sun_spin-current_2000}: 

\begin{equation}
 I_{s;crit} = \frac{4q}{\hbar}\alpha U \bigg(1+\frac{h_d}{2}\bigg)
 \label{eq:sun}
\end{equation}

However, in the case of GSHE-based switching employing PMA magnets introduces the well-known problem of indeterministic switching, since the polarization of the  injected spin-current from the GSHE is in the in-plane direction, thereby bringing the magnetization of the free layer to the in-plane hard axis, an unstable equilibrium position. A small magnetic field ($h_{bias} = H_{bias}/H_k$)  in the direction of the charge current helps break the symmetry of this equilibrium position and push the magnetization towards an easy axis. This field  may be provided either locally 
\cite{smith_external_2016} or as a small exchange-bias field built within the 
structure of the magnetic stack \cite{van_den_brink_field-free_2016}. 
We incorporate this field in our model simply as 
an additional external field provided to the LLG modules in the spin-circuit 
model.

In this case, the minimum spin-current necessary to switch the magnet is 
approximately given by  \cite{lee_threshold_2013}:

\begin{equation}
I_{s;crit} = \frac{4q}{\hbar} U \bigg(\frac{1}{2} - \frac{h_{bias}}{\sqrt{2}}\bigg)
\label{eq:hardaxis}
\end{equation}

It is interesting to note that the threshold current for ``hard-axis'' 
switching does not benefit from a  one to two orders of magnitude reduction due 
to the absence of the damping factor $\alpha$ that is present in 
eq.~\ref{eq:sun}. 

In fig.~\ref{fig_dynamic}b and fig.~\ref{fig_dynamic}c we observe that 
changing the magnetic layer in the  spin-switch from IMA to PMA reduces both 
the switching energy  
as well the static power dissipation, due to the relatively lower switching currents that allow
reduced supply voltages. 


\subsubsection{Magnet Design$-$High Anisotropy Scaled PMA}

Using PMA magnets in spin-switch opens up the possibility of using high 
anisotropy magnets commonly used in the magnetic recording 
industry. This is achieved by scaling down the grain volume ($\Omega$) and saturation magnetization 
($M_s$) while increasing the effective anisotropy  ($H_k$) to  maintain a given  
thermal stability since the barrier height is given by $U = M_s \Omega H_k/2$. While the minimum spin-current 
necessary to switch remains the same (eq.~\ref{eq:hardaxis}), the 
supply voltage levels  to produce the minimum switching current 
\textit{increase} as 
compared to unscaled PMA magnets due to two reasons: (a) increase in the 
resistance of the GSHE writer and (b) reduction of the geometrical gain due to a 
decrease in the length of the GSHE metal. 

However, using high $H_k$ and low $M_s \Omega$ reduce switching charge 
($Q_{sw}=2qM_s\Omega/\mu_B$) in 
the 
magnet, significantly improving the energy-delay as shown in 
fig.~\ref{fig_dynamic}b, since $E\times\tau=Q_{sw}^2 R$.
On the other hand, the reduction in $E\times\tau$ comes at a cost of increase 
in 
the static power (fig.~\ref{fig_dynamic}c) due to increased supply voltages, as 
shown in both red (medium scaled PMA) and blue (high scaled PMA) trajectories.

Increasing the $H_k$ of the magnets could necessitate changing 
the coupling mechanism between the WRITE and the READ units, since dipolar 
coupling may not provide strong enough interaction necessary for successful 
device operation. Advancements in magnetic oxides 
may open a pathway for stronger exchange-interaction based coupling of high 
$H_k$ magnets through either indirect exchange coupling mediated through 
the magnetic oxide \cite{sokalski_naturally_2013} or by replacing the 
metallic FM with an insulating FM \cite{li_spin-orbit_2016}. 
However, the thickness of the oxide layer necessary to provide exchange coupling 
is a critical issues that may interfere with the electrical isolation of the 
WRITE and the READ units through leakage currents due to tunneling effect, 
forming a parasitic MTJ between them. Our analysis presented here did not account for this 
leakage current in the coupling layer. Additionally, fabricating and creating 
contacts on highly scaled device could have lithographic challenges that may 
require change in the device design of the spin switch and is out of scope for 
our present study. Our focus here is to show how Modular Approach can be used 
to project performance enhancement assuming such challenges could be met.

\subsubsection{Reader Design$-$ Heusler Alloys}

A major improvement that can be incorporated in the spin-switch designs is the 
use of high polarization magnets (such as Heusler alloys) to fabricate 
MTJ or Spin-Valve based readers through an increase of $R_{OFF}/R_{ON}$. It can be 
shown from TMR formula ($R_{AP}-R_P / R_P = 2 P^2/1-P^2$) that to achieve the 
same 
$R_{OFF}/R_{ON}$ ratio as a well designed CMOS inverter ($\sim 10^5$), the 
interface polarization that would be required is $P = 0.99999$. For the sake of 
performance projection we choose an optimistic value of $P=0.99$ 
(close to an experimentally reported value of $P \approx 0.96$ at 
low temperatures \cite{liu2012giant}) which gives $R_{OFF}/R_{ON} \approx 
100$. 

Use of Heusler alloys helps in reducing the static power loss in the reader 
by reducing the supply voltages $V_{DD;SS}$ and bringing $V_{out}$ closer to the 
supply voltages (fig.~\ref{fig_dynamic}a). However, the high $P$ of the MTJ reader
imposes a limit on the overdrive that can be applied to the spin-switch, 
because the spin-current generated in the MTJ may start switching the device 
in competition with the GSHE writer, especially in scaled spin-switches where 
the geometrical gain in GSHE given by 
$\theta_{SH}{L}/{t}(1-\mathrm{sech}(t/\lambda_{sf}))$ 
is limited by scalability of GSHE thickness $t$ and spin-flip length 
$\lambda_{sf}$ compared to length $L$\cite{hong_spin_2016}. 

In fig.~\ref{fig_dynamic}b and fig.~\ref{fig_dynamic}c we see the effect of 
a reduction of both $E\times\tau$ and static power due to Heusler alloy MTJs in 
both red (medium scaled PMA) and green (ME spin-switch) trajectories. 

\subsubsection{Writer Design$-$High Spin Hall Angle}

Using materials with large spin Hall angles can help in scaled spin-switches by lowering the voltage levels,
since smaller charge currents can produce larger spin-current. For projection 
purposes, 
we choose a spin-Hall angle of $1$. In both blue (high scaled PMA) and red 
(medium scaled PMA) trajectories in 
fig.~\ref{fig_dynamic}b and fig.~\ref{fig_dynamic}c, the $E\times\tau$ is reduced 
by two orders of magnitude and static dissipation is reduced by one order of 
magnitude 
due to lower supply voltage levels.

We have assumed that the resistance of the GSHE material remains the same 
during this optimization, which is not necessarily true. In fact it has 
been suggested that resistance of most GSHE materials increase hand in hand with the Hall 
angle \cite{hong_spin_2016}, and this may reduce the magnitude of the 
$E\times\tau$ and 
static power improvements projected. However, it was recently suggested that 
using a composite structure consisting of a spin conduction layer between the 
GSHE and the magnet, it may be possible to obtain large spin-Hall angles from 
small spin-Hall angle materials \cite{shehrinpaper}. Such innovations in device 
engineering could provide a way to obtain large spin-Hall angles without the 
large resistance penalty.

\subsubsection{Writer Design$-$Spin Torque vs. Magnetoelectrics}
Instead of the spin-torque mechanism, voltage-controlled magnetoelectric (ME) 
phenomena based on multiferroic materials could be used for the WRITE in a 
spin-switch. In particular, 
$\rm BiFeO_3$ (BFO) was recently shown to be capable of switching an IMA magnet 
deterministically  \cite{heron_deterministic_2014} by 
applying a voltage controllable exchange bias field on the adjacent magnetic 
layer. 
Indeed, various device proposals  have used this mechanism as part of their 
device designs \cite{manipatruni_spin-orbit_2015,drcSamiran,drcMeghna}.

The switching process using ME fundamentally requires only $Q_{sw} = 
C_{ME}V$ 
amount of charge, where $C_{ME}$ is the capacitance of the ME material and $V$ 
is the applied voltage. Since the switching mechanism is not spin-torque, it can 
be 
much smaller compared to $Q_{sw} = 2qM_s\Omega/\mu_B$, opening a 
pathway for much more efficient 
switching. This phenomenon is also attractive because it creates a high input 
impedance device similar to a CMOS inverter and reduces static dissipation. 
Indeed, it is seen in fig.~\ref{fig_dynamic}b and fig.~\ref{fig_dynamic}c that 
the green trajectory that replacing the GSHE writer with an ME based writer 
reduces both $E\times\tau$ and static power drastically. 

 The ME module used in this work does not  consider the ferroelectric 
polarization caused by the electric field and hence 
is not a comprehensive model for the multiferroic material. Additionally, the dynamics 
of the ME based switching was deduced to be a complex 2-step process composed of 
two partial switchings in two different directions, ultimately causing a full 
reversal in the  experiment \cite{heron_deterministic_2014}. Coupling the 
LLG with an ME module produces only a single step switching. Since the detailed 
physics of multiferroic switching is not fully understood at this time, these 
projections are subject to change as a better understanding of voltage based 
multiferoic switching develops. 

\subsubsection{Magnetic Stack Design$-$ Synthetic Ferrimagnets}

Usage of synthetic ferrimagnet (Sy$-$AFM) stacks instead of monolayer magnets 
opens 
up an avenue of performance improvement of the spin-switch. This is due to a 
reduction in the effective switching charge ($Q_{sw} = 
2qM_s\Omega_{eff}/\mu_B$), where 
($M_s\Omega_{eff}=M_s\Omega_{1}-M_s\Omega_{2}$) compared to a  monodomain magnet 
($M_s\Omega_{total}=M_s\Omega_{1}+M_s\Omega_{2}$) with the same thermal 
stability, for details see \cite{camsari_ultrafast_2016}. In the blue 
trajectories (high scaled PMA) of fig.~\ref{fig_dynamic}b and 
fig.~\ref{fig_dynamic}c we use an Sy$-$AFM stack where  
$(M_s\Omega_{1}-M_s\Omega_{2})/(M_s\Omega_{1}+M_s\Omega_{2})\approx 1/3$. This 
provides an order of magnitude improvement in $E\times\tau$ alongside a 
\textit{reduction} of 
voltage levels by $40\%$ which helps reduce the static power. Optimized 
designs of Sy$-$AFM stacks may yield even higher performance gains.

\subsection{Outlook for the Spin-Switch}

Overall we find that spin-switches may approach the performance of 
contemporary scaled FinFET based CMOS if integration of various high 
performance materials along with careful device engineering 
and advanced lithographic and fabrication abilities can be 
achieved. Meanwhile, CMOS 
technology itself is a moving target, considering recent developments such as 
negative capacitance \cite{salahuddin2008use}, therefore it will be difficult 
for an 
individual spin device to  outperform the CMOS inverter in the near future.

Natural domain of spintronics may be in complex circuits where the inherent 
physics of a single device can map to a higher order logic function that 
requires many basic logic gates to implement, as argued in \cite{pan2016proposal}. These devices can then be deployed as compact and efficient computational nodes in complex Boolean and Beyond-Boolean architectures. The next sections explore this possibility using simple proof of concept demonstration.


\section{Beyond Boolean Circuits: Spintronic Probabilistic Networks}

A big thrust of spintronic research is the use of spin devices as ultra-compact 
\textit{deterministic} nodes of hardware neural networks 
\cite{sharad_spin-based_2012,sharad_spin-neurons:_2013,diep_spin_2014} due to 
inherent majority logic-like behavior (detailed analysis of such a compact 
majority logic circuit is provided in the supplementary). In this respect, the 
advantage of spin-switch as nodes of neural networks primarily lie in their 
ability to reduce dissipation and simpler circuit design due to reduced hardware 
cost compared to CMOS based implementations. In this section we go beyond 
the applications of spin-switch as neural network nodes and look at the physics 
of stochasticity of superparamagnets and its applications in the emerging field 
of probabilistic spintronic logic \cite{PSLpaper}. 
 
\subsection{Stochastic Magnet Dynamics}

Our spin-switch designs discussed up to this point have used magnets with  
$U = 40\ k_BT$ with state retention of nearly a decade. The state retention and 
barrier height is related by  $\tau_r = \tau_0 e^{U/k_BT}$, where $\tau_0 \sim 
0.1 - 1 \ \rm{ns}$. Even though we have not explicitly concerned
thermal noise in our analysis so far, when $U/k_BT \gg 1$, there is only a minor effect on the switching dynamics due to thermal agitation, and the switching delay is largely determined by the initial angle of the magnet. We have approximately taken this initial angle variation into consideration by using mean initial angles that are in consistent with results from  equilibrium statistical mechanics \cite{camsari_modular_2015-1}.

However, reducing the barrier height of these ferromagnets to superparagmagnetic values, for example $\approx3\ k_BT$, makes the 
 magnetization stochastic and the statistical average of the magnetization 
lies between ``up'' and ``down'' states. As an example, fig. \ref{fig_neural}a 
shows a transient simulation of a magnet with $U = 2.75\ k_BT$ only under the 
influence of a thermal noise field. The magnetization 
keeps flipping back and forth between 
the up and down states, since the state retention time is of the order of $\rm ns$. This 
stochastic behavior along with spin-torque or 
magnetic field switching can yield building blocks for probabilistic computers 
\cite{behin-aein_transynapse:_????,sutton2016intrinsic,PSLpaper} where probabilistic behavior comes 
naturally due to inherent physics of the device itself. 

\subsection{Stochastic Spin Switch: Building Block of Probabilistic Networks}

\begin{figure}[!t]
\centering
\includegraphics[width=2.8in]{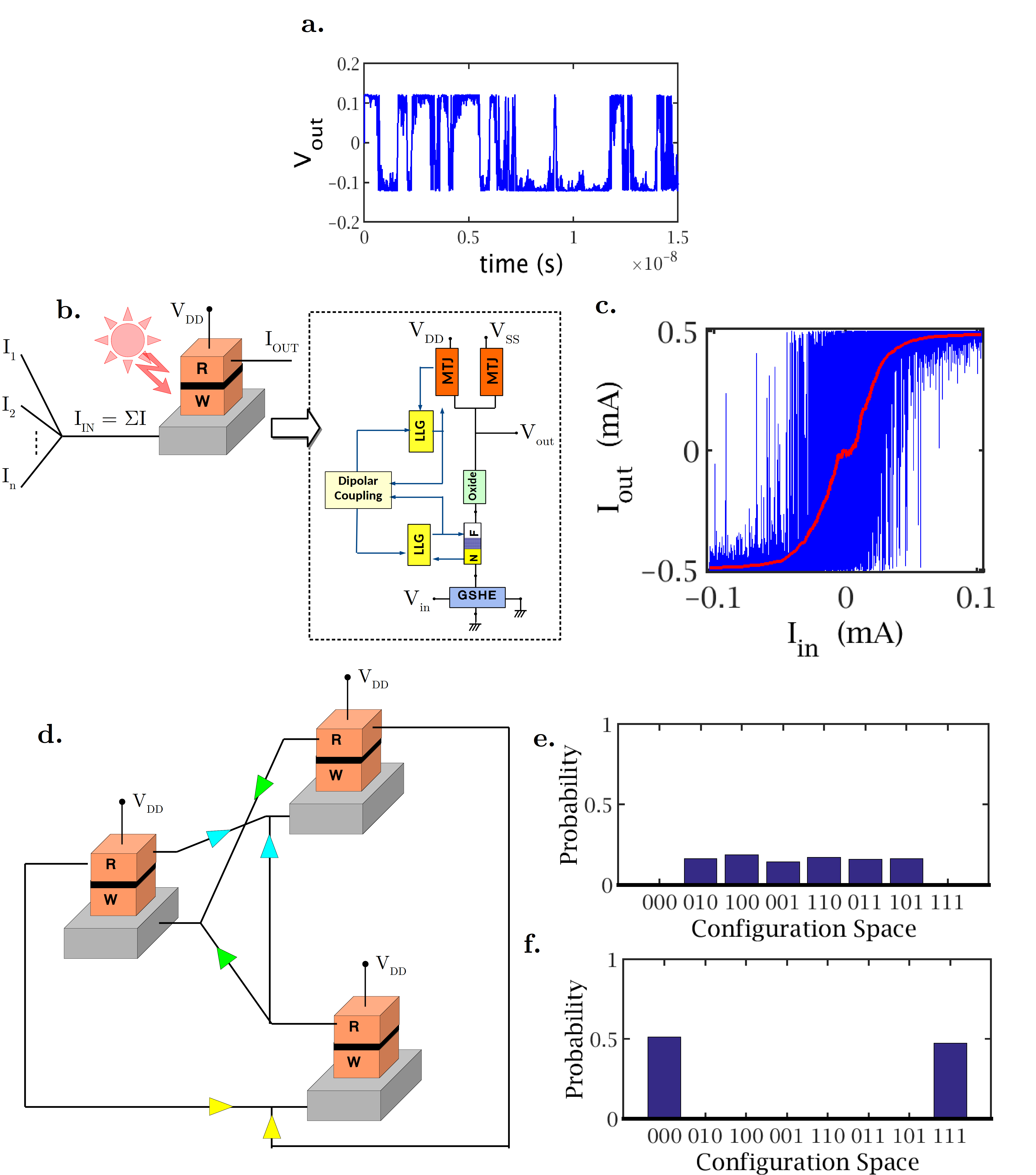}
\caption{\textbf{a. Stochastic Magnet Dynamics:} A low $U$ magnet keeps 
switching 
back and forth between ``up'' and ``down'' states when interacting with a 
thermal bath. This fluctuation can be read by looking at the output of 
the spin switch.
\textbf{b. Spin-Switch as a Stochastic Circuit Building Block:} Physics of spin-torque switching 
at the room temperature allows a spin-switch to behave as a building block for stochastic networks.
\textbf{c. Sigmoidal Transfer Characteristics of Stochastic Spin Switch:} A stochastic 
simulation showing the sigmoid function like transfer characteristics of the stochastic spin switch (blue background curve) whose mean magnetization can be changed by application of a magnetic field or, in this case, by a spin current. The time averaged mean (red foreground curve) can be obtained by passing the output of the spin switch through a low pass filter.
\textbf{d. Probabilistic Networks:} A three node Ising network built using 
the stochastic spin switch. Each node is driven by the other two nodes through 
the charge currents whose magnitude and sign can be programmed dynamically 
through supply voltages. \textbf{e.
Configuration Space $-$ Ferromagnetic:} Statistical sampling of the Ising 
network programmed for ferromagnetic type interaction shows that the network 
anneals to FM like states $000$ and $111$. \textbf{f. 
Configuration Space $-$ Frustrated Spin Glass:} Statistical sampling of the Ising network programmed for anti-ferromagnetic type interaction shows that the network anneals to frustrated spin-glass states. }
\label{fig_neural}
\end{figure}

Fig.~\ref{fig_neural}b shows a stochastic spin-switch operating under room temperature conditions. As explained 
in the previous section, the spin-switch naturally accumulates multiple input 
signals at its input node (analogous to synaptic addition in a neuron) and 
switches. When this switching happens at a finite temperature, the output of 
the neuron instead of being a deterministic function whose output is ``0'' or 
``1'', instead turns into a stochastic one whose average is skewed by the input 
current (blue background curve in fig.~\ref{fig_neural}c). This output when passed through 
a simple R-C low pass filter circuit ($\tau_{RC} = 225\ \rm{ns}$ in the simulation) 
that extracts the average value produces a transfer 
function which instead of being a sharp transition, turns into a sigmoidal 
function (red foreground curve in fig.~\ref{fig_neural}c), useful for building probabilistic and fuzzy logic circuits. 

The stochastic regime of operation of the spin-switch and its use as a building 
block for probablistic networks was first identified in 
\cite{behin-aein_transynapse:_????}, 
where it was called a \textit{transynapse}. 
The major distinction between the previous work and this paper is our use of 
spin-switches with in-plane magnets in the superparamagnetic regime and a time-averaged measurement, 
rather than averaging an ensemble of thermally stable magnets undergoing hard 
axis switching, following \cite{sutton2016intrinsic,camsari2016stochastic}. 
Since the computation with stochastic spin-switch is  statistical 
unlike deterministic switching, a direct comparison of  performance with 
another technology is highly implementation dependent and is not attempted 
here. 
Our main purpose is to demonstrate how the Modular Approach enables the exploration of 
stochastic spin-switches to build novel beyond-Boolean circuits.

\subsection{Programmable Stochastic Networks}

By connecting the stochastic spin switches together we can create novel 
beyond Boolean circuits, an example being the Ising network, a computational 
model that is widely used to solve complex optimization and pattern recognition 
problems \cite{z._bian_ising_2010}. The 
nodes of the network interact with each other through the charge currents 
in the GSHE, which can be controlled by either changing the 
voltage between the nodes (through sign and magnitude) or by using 
an external CMOS circuitry. The network can then be  annealed to its ground 
state providing a solution of the problem. While the Ising model in itself is purely 
deterministic, stochasticity of the superparamagnets helps  the system to traverse 
the configuration space of the network at the speed of the magnet retention time, 
which could be a few $\rm ns$ or less for superparamagnets.

As an example, fig.~\ref{fig_neural}e,f shows two different solutions 
mapped in the steady state statistical configuration of the 3-node Ising 
network (fig.~\ref{fig_neural}d), ferromagnetic and frustrated spin-glass. 
These two configurations can be obtained by tuning the interaction strength and 
sign between the three nodes. One way to achieve this is by an external 
circuitry that implements a weighing logic of the form $V_{in;i} = \sum w_{ij} 
V_{out;j}$ where $V_{in;i}$ is the input voltage for the $i^{th}$ node, 
$V_{out;j}$ is the output of the $j^{th}$ node and $w_{ij}$ is the weight 
logic and the elements of the Ising Hamiltonian for the given problem. In this 
problem $w_{ii} = 0$ and all positive $w_{ij}$ creates ferromagnetic 
interaction, whereas all negative $w_{ij}$ creates anti-ferromagnetic 
interaction. The magnitude of the $w_{ij} = \pm w_0$ is chosen to ensure that 
the average output of the nodes are in the ``saturated'' region of the 
fig.~\ref{fig_neural}a. The circuit is simulated in presence of magnetic 
thermal noise and the temperature of the system reduced slowly to anneal the 
circuit to its ground state. The time duration of these simulations were chosen 
to be three orders of magnitude higher compared to the state retention time of 
the superparamagnets to obtain stable statistics. The ref. 
\cite{sutton2016intrinsic} explains the simulation technique in more 
detail. 

If the interactions are tuned to make the network ferromagnetic, the 
configuration space obtained after a time averaged measurement is 
shown in fig.~\ref{fig_neural}f where nodes prefer the $000$ or 
$111$ state, i.e. all of the nodes are in the same state. However, for 
antiferromagnetic interaction, the nodes take up the other 6 possible 
states equally distributed statistically (fig.~\ref{fig_neural}f), forming a frustrated spin-glass, 
since the interaction strengths are equal in this simulation. 

This proof of concept demonstration points towards a possibility of building 
larger dynamically programmable stochastic networks (as in \cite{sutton2016intrinsic,PSLpaper}) that solve algorithms of 
data mining, optimization, searching, and machine learning (Deep Belief 
Networks) which at present are commercially implemented as 
software solutions. These stochastic networks map complex and composite 
logic functions directly into the physics of the spintronics, paving the way of 
creating ultra-compact and efficient learning networks.


\section{Conclusion}

In this paper, we have performed  
a  systematic analysis of a family of spin-based logic devices using the 
recently developed Modular Approach framework. By making materials, device and 
circuit level projections, we have estimated critical circuit metrics such as 
dynamic (WRITE unit) and static (READ unit) power dissipation of concrete 
designs and compared them to existing CMOS technology. With very 
optimistic material and design parameters that would only be available with 
several technological breakthroughs in the field, such as Heusler alloys and 
high spin Hall angles, we find that spin-based logic devices could have 
energy-delay products
comparable to CMOS technology. Besides there are factors such as non-volatility 
that are not captured by the energy-delay product.

It is important to point out however, given the enormous rate of discoveries in the field, it is very likely that the conclusions presented in this paper will quickly need to be updated, and as such, one of the key objectives of this paper was to illustrate how the Modular Approach framework can reliably integrate emerging physics into existing physics at the device and circuit level. 

As an example of the versatility of the modular approach, we show that 
even for beyond-Boolean
architectures different from what is conventionally discussed in spin-logic proposals,
the same framework allows systems-level analyses, once an experimentally benchmarked module
for the relevant building blocks (superparamagnets in this case) is available.

\section*{Acknowledgment}

The authors would like to extend their thanks to Dr. Behtash Behin-Aein for 
sharing his preprint \cite{behin-aein_transynapse:_????}  and for helpful 
discussions on programmable stochastic networks. The authors would also 
like to thank Dr. Azad Naeemi for helpful discussions on engineering 
challenges with the spin-switch and the alternative designs. This 
work 
was supported in part by the National Science Foundation through the 
NCN-NEEDS program, contract 1227020-EEC and in part by C-SPIN, one of six 
centers of STARnet, a Semiconductor Research Corporation program, sponsored by 
MARCO and DARPA.




\bibliographystyle{IEEEtran}
\bibliography{spinpower}

%

%








\end{document}


\maketitle

\section{Device Equations for the Spin Switch family}

For the spin switch with grounded GSHE layer, the minimum logic voltage 
level ($V_{1,2,\cdots} = V_{LL}$) and supply voltage levels ($V_{DD},V_{SS}$) 
necessary for switching are given by the following set of equations. 

Basic fundamental material properties:

\begin{eqnarray}
& 2U =  M_s\Omega H_k \\
& g_{MTJ;P} = g_{MTJ;0}(1 + P_1P_2 \hat M\cdot \hat m_{r}(t)) \\
& g_{MTJ;AP} =  g_{MTJ;0}(1 - P_1P_2 \hat M\cdot \hat m_{r}(t)) \\
& g_{GSHE} =  \rho_{GSHE} \displaystyle{\frac{L_{GSHE}}{W_{GSHE}\tau_{GSHE}}} \\
& g_{FM-NM} =  \displaystyle{\Re(g_{\uparrow\downarrow})A_{FM}}
\end{eqnarray}

Critical spin current for IMA magnet:

\begin{equation}
 I_{s;crit} = 4q\alpha U(1+\frac{2\pi M_s}{H_k})
\end{equation}

Critical spin current for PMA magnet:

\begin{equation}
 I_{s;crit} = 4qU(\frac{1}{2}-\frac{H_{w;ext.}}{\sqrt{2}H_k})
\end{equation}

Transport relationship between GSHE charge current and critical spin current 
for switching the magnet:

\begin{equation}
k_1 = 
\theta_{sh}\frac{A_{FM}}{W\tau}\frac{\frac{g_{FM-NM}}{
g_{GSHE}}}{ csch(\frac{\tau}{\lambda_{sf}}) + 
tanh(\frac{\tau}{2\lambda_{sf}}) 
+\frac{g_{FM-NM}}{g_{GSHE}}}g_{GSHE}
\end{equation}

\noindent where $W,\tau$ are width and dimensions of the GSHE material.

Voltage divider equation between the READ stage of the spin switch and the next 
stage (load):

\begin{equation}
k_2 = \displaystyle{\frac{g_{MTJ;P}-g_{MTJ;AP}}{g_{MTJ;P}+g_{MTJ;AP}+g_{load}}}
\end{equation}

Relationship between the logic voltage level and the critical spin current:

\begin{equation}
V_{LL} = \frac{I_{s;crit}}{k_1}
\end{equation}

From the constitutive relationship between the electric and magnetic fields in 
the ME:

\begin{equation}
\alpha_{ME} = \mu_0 \frac{dM}{dE}
\end{equation}

We can derive the relationship between logic voltage level and critical 
switching field for ME-FM system in ME spin switch (the threshold switching 
field of a magnet is $H_{sw} = H_k$, in a monodomain approximation) is :

\begin{equation}
V_{LL} = \frac{H_k t_{ME}}{\alpha_{ME}}
\end{equation}

Relationship between the logic voltage level and supply voltage levels:

\begin{equation}
V_{DD;SS} = (\pm)\frac{V_{LL}}{k_2} 
\end{equation}

The symbols are noted in parameters section, $V_{LL}$ is the logic voltage 
level and is not necessarily the same as the supply voltages $V_{DD;SS}$. The 
simulation performed for fig. 2 (main paper) are at near minimal overdrive to 
be close to the numerically conditions described by the analytical expressions 
at switching threshold.

The $g_{load}$, in the device equations stands for the output 
loading on the device and can be substituted (for an FO-1 circuit) with  
$g_{GSHE}$ as a 
reasonable approximation of the input admittance of the next stage  due to the GSHE 
material. We ignore the spin-Hall magnetoresistance effect (SMR effect) \cite{chen_theory_2013} due to the 
write magnet of the next stage and its dynamics while switching. The SMR effect 
is  automatically incorporated in the numerical spin-circuit model  through the GSHE 
module but does not introduce a significant discrepancy from 
the above analytical expressions at steady state since it is a second-order
effect proportional to $\theta^2_{SH}\ll 1$. For ME based spin switch the 
$g_{load} = 0$ at steady state because a charged capacitor behaves as an open 
circuit.

\section{Operating Points for the Spin Switch family}
 
From the device equations above, it is possible to calculate the minimum 
operating point currents and voltages for the spin switch. The 
table below lists these operating points for the devices discussed in 
section III of the main paper. These points are calculated from the equations 
above and use the parameters provided in the last section of the supplementary.

\medskip
\begin{center}
\begin{tabular}{|l|l|l|l|}
\hline
\multirow{2}{*}{\textbf{Device}} & \textbf{$\rm I_{s;crit}$} & 
\textbf{$\rm V_{LL}$} & \textbf{$\rm V_{DD}$} \\ 
& \textbf{$\rm (\mu A)$} & \textbf{$\rm (mV)$} & \textbf{$\rm (mV)$} \\
\hline
IMA Spin Switch (SS) & $401.9$ & $22.7$ & $80$ \\
Low $H_k$ PMA SS & $235.8$ & $17$ & $55$ \\
PMA SS & $189.1$ & $42.6$ & $138.2$ \\
High $H_k$ Scaled PMA SS & $196.7$ & $88.6$ & $287.3$ \\
High SHA high $H_k$ PMA SS & $196.7$ & $26.6$ & $86.1$ \\
SyAFM SS & $65.4$ & $8.8$ & $28.7$ \\
High SHA PMA SS & $189.1$ & $12.8$ & $41.5$ \\
Heusler MTJ PMA SS & $189.1$ & $12.8$ & $20.7$ \\
ME SS & - & $13$ & $26.5$ \\
Heusler MTJ ME SS & - & $13$ & $13.2$ \\
High ME SS & - & $433.3$ & $442.2$ \\
\hline
\end{tabular}
\end{center}
\medskip

\section{Spin-Circuit Modules}
 
The modular spintronic library of the consists of two kinds of modules.

\textbf{Transport Modules:} These modules capture the charge and spin 
transport through various materials, e.g. Normal metal (NM), Ferromagnet (FM), 
FM$-$NM Interface, Magnetic Tunnel Junctions (MTJ) and various spin-orbit 
materials such as Giant Spin Hall Effect (GSHE).

\textbf{Magnetics Modules:} These modules capture the physics of 
magnetic dynamics and interaction and are meant to be coupled with the 
transport modules. The modules available are: Landau-Lifshitz-Gilbert Equation 
Solver (LLG), Exchange Coupling, Dipolar Coupling, and Magneto-Electric Effect 
(ME).
 
Most of the Spin-Circuit modules used in this work have been described in the 
previous work \cite{camsari_modular_2015} and their source code can be obtained 
from the online repository for the project \cite{_modular_web}. In this work we 
introduced the new magnetoelectric module that we briefly describe below.

\subsection{Magneto-Electric Module}
The magneto-electric effect is given by:
\begin{equation}
 \alpha_{ME} = \mu_0\frac{dM}{dE}
\end{equation}
Now:
\begin{equation}
B = \mu_0 M 
\end{equation}
Therefore, 
\begin{equation}
 dB_{ME} = \alpha_{ME}dE_{ME}
\end{equation}

\noindent Where $E_{ME} = V_{ME}/t_{ME}$ is the electric field on the multi-ferroic 
material, $B_{ME}$ is the generated exchange field on an adjacent magnetic 
layer, and $\alpha_{ME}$ is the empirically measured coefficient for the effect.

\begin{figure}[!h]
\centering
\includegraphics[width=2.5in]{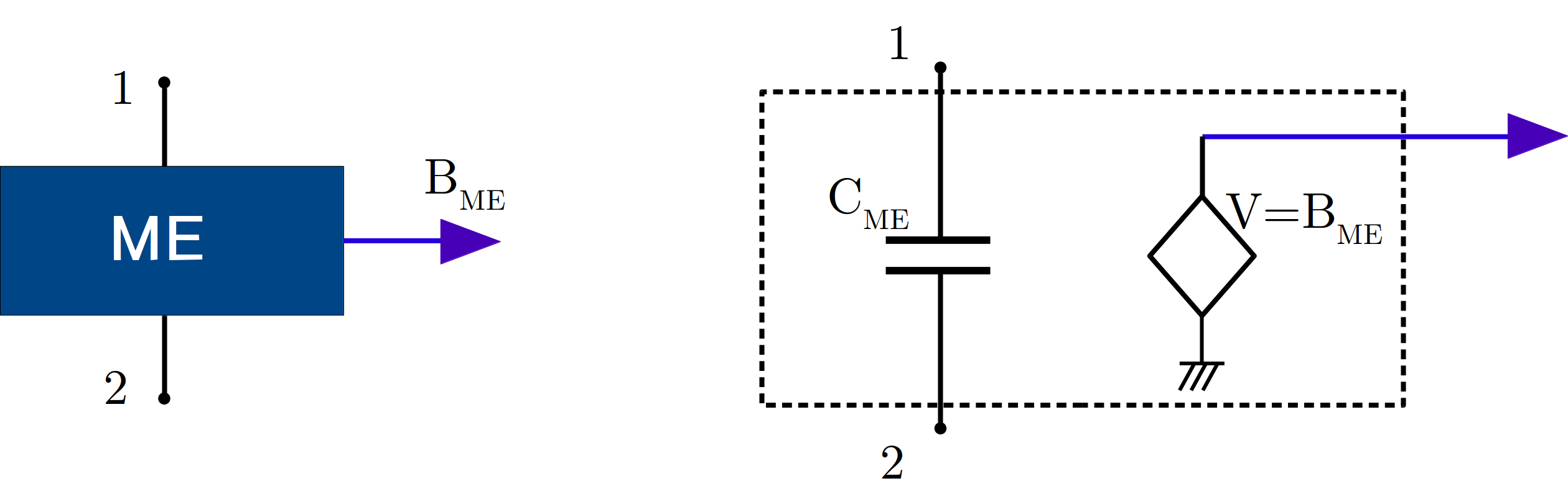}
\caption{Spin-Circuit model for Magneto-Electric module}
\label{fig_memodule}
\end{figure}

The effect is modeled as a parallel plate capacitor and a controlled voltage 
source whose strength is the magnetic field produced by the ME layer as shown in 
fig.~\ref{fig_memodule}. 

\medskip
\begin{center}
\begin{tabular}{|l|l|l|}
\hline
\textbf{Parameter} & \textbf{Symbol} & \textbf{Units} \\ \hline
Area & $A_{ME}$ & $\rm m^2$ \\
Thickness & $t_{ME}$ & $\rm m$ \\
ME Coefficient & $\alpha_{ME}$ & $\rm s/m $ \\ 
Relative Permittivity & $\epsilon_r$ & $-$ \\ \hline
\end{tabular}
\end{center}
\medskip


\section{Complex Boolean Functions: Majority Logic}

Spintronic phenomena enables many higher order Boolean functions to be 
implemented using minimal hardware compared to CMOS devices. An 
example of this is the majority function. In this section we use the Modular 
Approach to investigate the physics behind this capability. 

\begin{figure}[!t]
\centering
\includegraphics[width=3.4in]{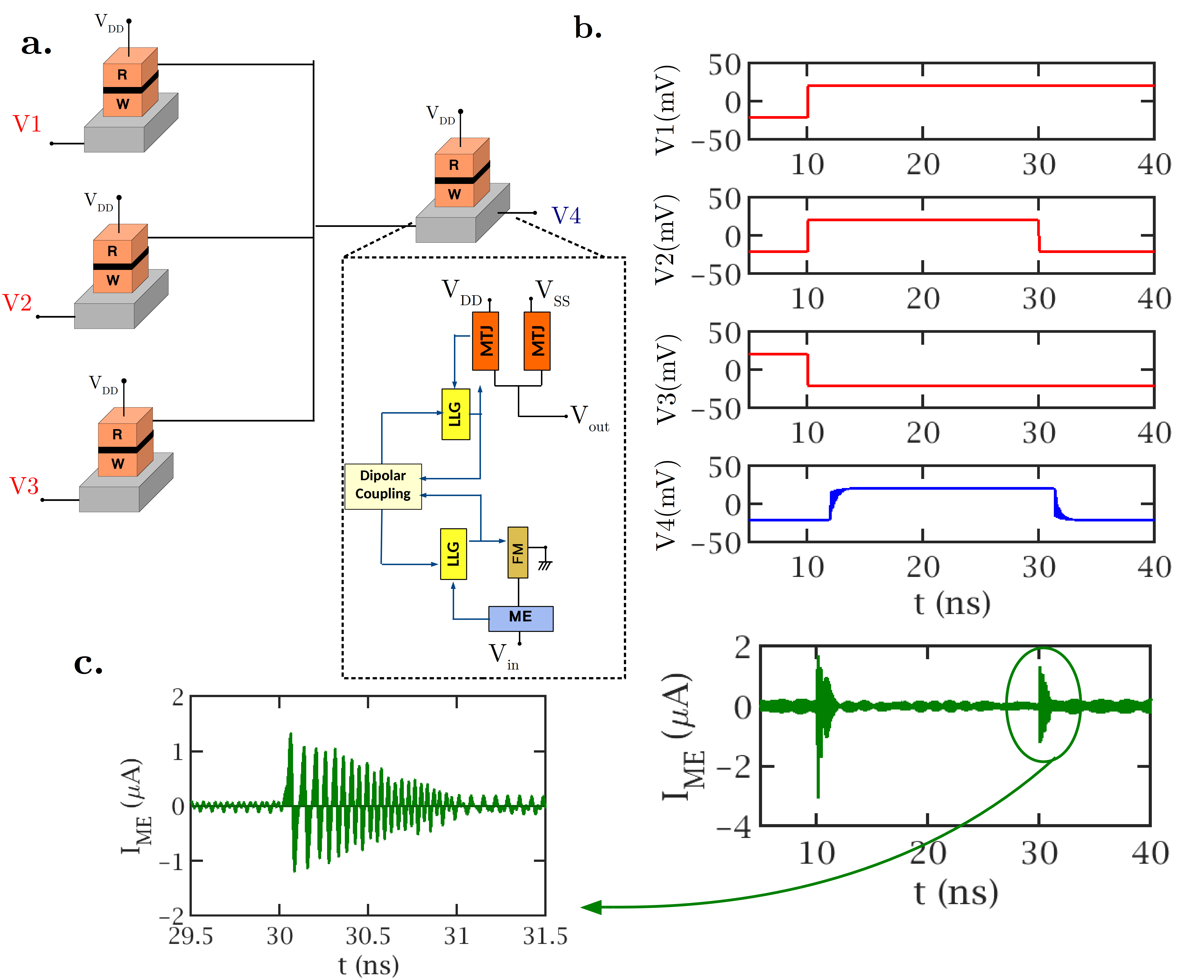}
\caption{\textbf{a. Majority Gate Circuit using Spin Switch:} A 3-input majority logic circuit using the ME 
Spin-Switch \textbf{b. Demonstration of Majority Function:} A simulation showing the majority function realization. 
The output is $V4 = Majority(V1, V2, V3)$ at steady state of the circuit. \textbf{c. Switching Charge $\mathbf{Q_{sw}}$ for ME Spin Switch:} Net input current into the output switch for the second switching event. Integral of the current in the 
switching window $>$ threshold charge on ME capacitor for switching, and is comparable to a CMOS inverter's $Q_{sw}$.}
\label{fig_majority}
\end{figure}

In fig. 2f (in the main paper) we showed that critical charge necessary to switch 
a magnet can be related to the number of spins $2M_s\Omega/\mu_B$ comprising 
the magnet, and in the case of ME writer, it is the threshold charge 
(eq.~\ref{eq:me}) on the ME capacitor. This charge can be provided through 
currents coming in from multiple inputs all adding in without any extra 
circuitry, since it is more natural to add currents in metal interconnects 
unlike adding voltages. This fan-in capability of spintronic devices leads to 
the realization that higher level logic functions, such as majority function 
based logic are a natural domain of spintronics 
\cite{srinivasan_all-spin_2011,nikonov_proposal_2011,diep_spin_2014}. 

\subsection{Majority Gate using the Magnetoelectric Spin-Switch} 

We investigate the physics of a majority function implemented using a single 
ME spin-switch and show that this may open the pathway for low dissipation 
complex logic circuits compared to those built using larger number of basic 
logic gates implemented through CMOS inverters \cite{smolensky_algebraic_1987}. 
Fig.~\ref{fig_majority}a shows a 3 input majority function being implemented 
using an ME spin-switch. The output V4 implements the Majority function 
over the input signals V1, V2, V3, as can be seen from the simulation in 
fig.~\ref{fig_majority}b. In the last plot of the fig.~\ref{fig_majority}b we 
plot the current in to the output switch, and the fig.~\ref{fig_majority}c 
shows the current zoomed in for the second transition event. 

\subsection{Dissipative Cost of ME-SS Majority Function} 

It can be shown that the minimum charge necessary to switch the 
magnet is given by:

\begin{equation}
 Q_{crit} = \epsilon_{0}\epsilon_{r}\frac{A_{ME}H_k}{\alpha_{ME}}
 \label{eq:me}
\end{equation}

For the chosen parameters the critical charge to switch the output 
device is $Q_{crit} = 100\ e^-$. Numerically we find that in the 
switching window (for either of the switching events): $Q_{ME} = \int I_{ME} 
dt = \int I_{sw1} dt + \int I_{sw2} dt + \int I_{sw3} dt \approx 190\ e^- $, 
which is close to the theoretical limit.

This number demonstrates that for the dissipative cost of just one CMOS  
inverter, ME spin-switch can implement a majority function as a basis of more 
complex functions, such as full adders that need a large number of CMOS  
inverters to implement \cite{moaiyeri_new_2009}, or neural networks 
\cite{diep_spin_2014}. 







\section{Numerical Parameters For Spin-Circuit Models}

We list the baseline parameters for various modules used in this work below. In 
simulation only one parameter is changed at a time, except for magnet design 
where the area of the magnet or $M_s$ is changed with $H_k$ 
to maintain an energy barrier of $40 \ kT$. All parametric changes are monotonic. Note: $n$ stands for nano and has been used to reduce visual clutter from the tables.
\begin{center}
\textit{MTJ Module}
\end{center}
\begin{center}
\begin{tabular}{|l|l|l|l|}
\hline
\textbf{Parameter} & \textbf{Units} & \textbf{Value} \\ \hline
Conductance & $\rm S$ & $1m$ \\
Polarization & - & $0.7$  \\
In-Plane Coeff. & - & $1$ \\ 
OOP Coeff. & - & $0$ \\ \hline
\end{tabular}
\end{center}
\medskip

\begin{center}
\textit{MTJ Module: Heusler Alloy}
\end{center}
\begin{center}
\begin{tabular}{|l|l|l|l|}
\hline
\textbf{Parameter} & \textbf{Units} & \textbf{Value} \\ \hline
Conductance & $\rm S$ & $1m$ \\
Polarization & - & $0.99$  \\
In-Plane Coeff. & - & $1$ \\ 
OOP Coeff. & - & $0$ \\ \hline
\end{tabular}
\end{center}
\medskip
\begin{center}
\textit{MTJ Module: Heusler Alloy SV}
\end{center}
\begin{center}
\begin{tabular}{|l|l|l|l|}
\hline
\textbf{Parameter} & \textbf{Units} & \textbf{Value} \\ \hline
Conductance & $\rm S$ & $0.1$ \\
Polarization & - & $0.99$  \\
In-Plane Coeff. & - & $1$ \\ 
OOP Coeff. & - & $0$ \\ \hline
\end{tabular}
\end{center}
\medskip
\begin{center}
\textit{GSHE Module: IMA Magnet}
\end{center}
\begin{center}
\begin{tabular}{|l|l|l|l|}
\hline
\textbf{Parameter} & \textbf{Units} & \textbf{Value} \\ \hline
Width & $\rm m$ & $100n$\\
Length & $\rm m$ & $80n$\\
Thickness & $\rm m$ & $2n$\\
Spin-flip length & $\rm m$ & $2n$\\
Resistivity & $\rm \Omega  m$ & $1700 n$ \\ 
Spin Hall Angle & - & $0.3$ \\ \hline
\end{tabular}
\end{center}
\medskip
\begin{center}{\textit{GSHE Module: Low $H_k$ PMA Magnet}}
\end{center}
\begin{center}
\begin{tabular}{|l|l|l|l|}
\hline
\textbf{Parameter} & \textbf{Units} & \textbf{Value} \\ \hline
Width & $\rm m$ & $100n$\\
Length & $\rm m$ & $100n$\\
Thickness & $\rm m$ & $2n$\\
Spin-flip length & $\rm m$ & $2n$\\
Resistivity & $\rm \Omega  m$ & $1700 n$ \\  
Spin Hall Angle & - & $0.3$ \\ \hline
\end{tabular}
\end{center}
\medskip
\begin{center}{\textit{GSHE Module: Medium $H_k$ PMA Magnet}}
\end{center}
\begin{center}
\begin{tabular}{|l|l|l|l|}
\hline
\textbf{Parameter} & \textbf{Units} & \textbf{Value} \\ \hline
Width & $\rm m$ & $32n$\\
Length & $\rm m$ & $32n$\\
Thickness & $\rm m$ & $2n$\\
Spin-flip length & $\rm m$ & $2n$\\
Resistivity & $\rm \Omega  m$ & $1700 n$ \\ 
Spin Hall Angle & - & $0.3$ \\ \hline
\end{tabular}
\end{center}
\medskip
\begin{center}
\textit{GSHE Module: High $H_k$ PMA Magnet}
\end{center}
\begin{center}
\begin{tabular}{|l|l|l|l|}
\hline
\textbf{Parameter} & \textbf{Units} & \textbf{Value} \\ \hline
Width & $\rm m$ & $16n$\\
Length & $\rm m$ & $16n$\\
Thickness & $\rm m$ & $2n$\\
Spin-flip length & $\rm m$ & $2n$\\
Resistivity & $\rm \Omega  m$ & $1700 n$ \\ 
Spin Hall Angle & - & $0.3$ \\ \hline
\end{tabular}
\end{center}
\medskip
\begin{center}
\textit{FM-NM Module: IMA Magnet}
\end{center}
\begin{center}
\begin{tabular}{|l|l|l|l|}
\hline
\textbf{Parameter} & \textbf{Units} & \textbf{Value} \\ \hline
Conductance & $\rm S$ & $80n \times 100n \times 5\times10^{15}$ \\
Polarization & - & $0.7$  \\
In-Plane Coeff. & - & $1$ \\ 
OOP Coeff. & - & $0$ \\ \hline
\end{tabular}
\end{center}
\medskip

\begin{center}
\textit{FM-NM Module: Low $H_k$ PMA Magnet}
\end{center}

\begin{center}
\begin{tabular}{|l|l|l|l|}
\hline
\textbf{Parameter} & \textbf{Units} & \textbf{Value} \\ \hline
Conductance & $\rm S$ & $\pi \times 50n \times 50n \times 5\times10^{15}$ \\
Polarization & - & $0.7$  \\
In-Plane Coeff. & - & $1$ \\ 
OOP Coeff. & - & $0$ \\ \hline
\end{tabular}
\end{center}

\begin{center}\textit{FM-NM Module: Medium $H_k$ PMA Magnet}
\end{center}
\begin{center}
\begin{tabular}{|l|l|l|l|}
\hline
\textbf{Parameter} & \textbf{Units} & \textbf{Value} \\ \hline
Conductance & $\rm S$ & $\pi \times 16n \times 16n \times 5\times10^{15}$ \\
Polarization & - & $0.7$  \\
In-Plane Coeff. & - & $1$ \\ 
OOP Coeff. & - & $0$ \\ \hline
\end{tabular}
\end{center}
\begin{center}
\textit{FM-NM Module: High $H_k$ PMA Magnet}
\end{center}
\begin{center}
\begin{tabular}{|l|l|l|l|}
\hline
\textbf{Parameter} & \textbf{Units} & \textbf{Value} \\ \hline
Conductance & $\rm S$ & $\pi \times 8n \times 8n \times 5\times10^{15}$ \\
Polarization & - & $0.7$  \\
In-Plane Coeff. & - & $1$ \\ 
OOP Coeff. & - & $0$ \\ \hline
\end{tabular}
\end{center}
\begin{center}
\textit{LLG Module: IMA Magnet}
\end{center}
\begin{center}
\begin{tabular}{|l|l|l|l|}
\hline
\textbf{Parameter} & \textbf{Units} & \textbf{Value} \\ \hline
Area & $\rm m^2$ & $80n \times 100n$ \\
Thickness & $\rm m$ & $2n$  \\
Damping Coeff. & - & $0.01$ \\ 
Sat. Magnetization & emu/cc & $800$ \\ 
Aniso. Field Strength & $\rm Oe$ & $130$ \\ \hline
\end{tabular}
\end{center}

\begin{center}
\textit{LLG Module: Low $H_k$ PMA Magnet}
\end{center}
\begin{center}
\begin{tabular}{|l|l|l|l|}
\hline
\textbf{Parameter} & \textbf{Units} & \textbf{Value} \\ \hline
Area & $\rm m^2$ & $\pi\times50n\times50n$ \\
Thickness & $\rm m$ & $2n$  \\
Damping Coeff. & - & $0.01$ \\ 
Sat. Magnetization & emu/cc & $800$ \\ 
Aniso. Field Strength & $\rm Oe$ & $130$ \\ \hline
\end{tabular}
\end{center}
\begin{center}
\textit{LLG Module: Medium $H_k$ PMA Magnet}
\end{center}
\begin{center}
\begin{tabular}{|l|l|l|l|}
\hline
\textbf{Parameter} & \textbf{Units} & \textbf{Value} \\ \hline
Area & $\rm m^2$ & $\pi\times 16n\times 16n$ \\
Thickness & $\rm m$ & $2n$  \\
Damping Coeff. & - & $0.1$ \\ 
Sat. Magnetization & emu/cc & $400$ \\ 
Aniso. Field Strength & $\rm Oe$ & $2500$ \\ \hline
\end{tabular}
\end{center}
\begin{center}
\textit{LLG Module: High $H_k$ PMA Magnet}
\end{center}
\begin{center}
\begin{tabular}{|l|l|l|l|}
\hline
\textbf{Parameter} & \textbf{Units} & \textbf{Value} \\ \hline
Area & $\rm m^2$ & $\pi\times 8n\times 8n$ \\
Thickness & $\rm m$ & $2n$  \\
Damping Coeff. & - & $0.1$ \\ 
Sat. Magnetization & emu/cc & $400$ \\ 
Aniso. Field Strength & $\rm Oe$ & $10100$ \\ \hline
\end{tabular}
\end{center}
\medskip
\begin{center}
\textit{LLG Module: High $H_k$ PMA Sy$-$AFM}
\end{center}
\begin{center}
\begin{tabular}{|l|l|l|l|}
\hline
\textbf{Parameter} & \textbf{Units} & \textbf{Value} \\ \hline
Area & $\rm m^2$ & $\pi\times 8n\times 8n$ \\
Thickness Assist & $\rm m$ & $1.2n$  \\
Thickness Free & $\rm m$ & $0.8n$  \\
Damping Coeff. & - & $0.1$ \\ 
Sat. Magnetization & emu/cc & $400$ \\ 
Aniso. Field Strength & $\rm Oe$ & $10100$ \\ \hline
\end{tabular}
\end{center}
\medskip
\begin{center}
\textit{LLG Module: High $H_k$ IMA}
\end{center}
\begin{center}
\begin{tabular}{|l|l|l|l|}
\hline
\textbf{Parameter} & \textbf{Units} & \textbf{Value} \\ \hline
Area & $\rm m^2$ & $ 8n\times 20n$ \\
Thickness & $\rm m$ & $2n$  \\
Damping Coeff. & - & $0.05$ \\ 
Sat. Magnetization & emu/cc & $400$ \\ 
Aniso. Field Strength & $\rm Oe$ & $13000$ \\ \hline
\end{tabular}
\end{center}
\medskip
\begin{center}
\textit{ME Module}
\end{center}
\begin{center}
\begin{tabular}{|l|l|l|l|}
\hline
\textbf{Parameter} & \textbf{Units} & \textbf{Value} \\ \hline
Area & $\rm m^2$ & $80n \times 100n$ \\
Thickness & $\rm m$ & $10n$  \\
ME Coeff. & - & $1\times10^{-8}$ \\ 
Rel. Permittivity & - & $500$ \\  \hline
\end{tabular}
\end{center}
\medskip
\begin{center}
\textit{High ME Module}
\end{center}
\begin{center}
\begin{tabular}{|l|l|l|l|}
\hline
\textbf{Parameter} & \textbf{Units} & \textbf{Value} \\ \hline
Area & $\rm m^2$ & $80n \times 100n$ \\
Thickness & $\rm m$ & $10n$  \\
ME Coeff. & - & $3\times10^{-8}$ \\ 
Rel. Permittivity & - & $500$ \\  \hline
\end{tabular}
\end{center}
\medskip
\begin{center}
\textit{Dipolar Coupling Module: IMA Magnet}
\end{center}
\begin{center}
\begin{tabular}{|l|l|l|l|}
\hline
\textbf{Parameter} &  \textbf{Units} & \textbf{Value} \\ \hline
Sat. Magn. 1 &  $\rm emu/cc$ & $800$ \\
Vol. 1 &  $\rm m^3$ & $100n\times80n\times2n$  \\
Sat. Magn. 2 &  $\rm emu/cc$ & $800$ \\
Vol. 2 & $\rm m^3$ & $100n\times80n\times2n$  \\
Dipolar Coeff. &  - & $0.0256,-0.0128,-0.0128$ \\ \hline
\end{tabular}
\end{center}
\medskip
\begin{center}
\textit{Exchange Coupling Module: Low $H_k$ PMA Magnet}
\end{center}
\begin{center}
\begin{tabular}{|l|l|l|l|}
\hline
\textbf{Parameter} &  \textbf{Units} & \textbf{Value} \\ \hline
Sat. Magn. 1 & $\rm emu/cc$ & $800$ \\
Vol 1 & $\rm m^3$ & $\pi\times50n\times50n\times2n$  \\
Sat. Magn. 2 & $\rm emu/cc$ & $800$ \\
Vol 2 & $\rm m^3$ & $\pi\times50n\times50n\times2n$  \\
Exchange Field Coeff. & $\rm erg/cm^2$ & $5$ \\ \hline
\end{tabular}
\end{center}
\medskip
\begin{center}
\textit{Exchange Coupling Module: Medium $H_k$ PMA Magnet}
\end{center}
\begin{center}
\begin{tabular}{|l|l|l|l|}
\hline
\textbf{Parameter} &  \textbf{Units} & \textbf{Value} \\ \hline
Sat. Magn. 1 & $\rm emu/cc$ & $400$ \\
Vol 1 & $\rm m^3$ & $\pi\times16n\times16n\times2n$  \\
Sat. Magn. 2 & $\rm emu/cc$ & $400$ \\
Vol 2 & $\rm m^3$ & $\pi\times16n\times16n\times2n$  \\
Exchange Field Coeff. & $\rm erg/cm^2$ & $5$ \\ \hline
\end{tabular}
\end{center}
\medskip
\begin{center}
\textit{Exchange Coupling Module: High $H_k$ PMA Magnet}
\end{center}
\begin{center}
\begin{tabular}{|l|l|l|l|}
\hline
\textbf{Parameter} &  \textbf{Units} & \textbf{Value} \\ \hline
Sat. Magn. 1 & $\rm emu/cc$ & $400$ \\
Vol 1 & $\rm m^3$ & $\pi\times8n\times8n\times2n$  \\
Sat. Magn. 2 & $\rm emu/cc$ & $400$ \\
Vol 2 & $\rm m^3$ & $\pi\times8n\times8n\times2n$  \\
Exchange Field Coeff. & $\rm erg/cm^2$ & $5$ \\ \hline
\end{tabular}
\end{center}
\medskip
\begin{center}
\textit{Exchange Coupling Module: Sy$-$AFM}
\end{center}
\begin{center}
\begin{tabular}{|l|l|l|l|}
\hline
\textbf{Parameter} &  \textbf{Units} & \textbf{Value} \\ \hline
Sat. Magn. 1 & $\rm emu/cc$ & $400$ \\
Vol Assist & $\rm m^3$ & $\pi\times8n\times8n\times1.2n$  \\
Sat. Magn. 2 & $\rm emu/cc$ & $400$ \\
Vol Free & $\rm m^3$ & $\pi\times8n\times8n\times0.8n$  \\
Exchange Field Coeff. & $\rm erg/cm^2$ & $5$ \\ \hline
\end{tabular}
\end{center}
\medskip
\begin{center}
{\textit{LLG Module: Stochastic Magnet}}
\end{center}
\begin{center}
\begin{tabular}{|l|l|l|l|}
\hline
\textbf{Parameter} & \textbf{Units} & \textbf{Value} \\ \hline
Area & $\rm m^2$ & $30n \times 15n$ \\
Thickness & $\rm m$ & $0.5n$  \\
Damping Coeff. & - & $0.01$ \\ 
Sat. Magn. & emu/cc & $500$ \\ 
Aniso. Field Strength & $\rm Oe$ & $1000$ \\ \hline
\end{tabular}
\end{center}

The FM-NM interface conductance values in our modules correspond to half 
the magnitudes reported in the literature, for instance see
\cite{brataas_non-collinear_2006,chen_theory_2013}. 
$g_0 = 5\times10^{15}$ corresponds to the $Re(g_{\uparrow\downarrow}) = 
2.5\times10^{15}$.

\bibliographystyle{IEEEtran}
\bibliography{spinpower}